\documentclass[11pt]{article}
\usepackage{jheppub,amsmath, amsthm, amssymb,slashed,url}
\usepackage{graphicx}
\usepackage{epstopdf}
\usepackage{tikz}
\usepackage{xcolor}
\usepackage{tabu}
\usepackage{tabularx}

\usepackage{etoolbox}
    \makeatletter
    \patchcmd{\maketitle}{\@fpheader}{}{}{}
    \makeatother

\usepackage{boldline}
\usepackage{bm}
\usepackage{amssymb,amsfonts}
    
\def\oneht{\textstyle{1\over 2} }

\def\OMIT#1{{}}

\def\si{^1 \hskip -0.03in S _0}
\def\siii{^3 \hskip -0.025in S _1}

\newcommand{\beq}{\begin{equation}}
\newcommand{\eeq}{\end{equation}}
\newcommand{\beqa}{\begin{eqnarray}}
\newcommand{\eeqa}{\end{eqnarray}}

\newcommand{\nn}{\nonumber}

\dedicated{NT@UW-21-15, IQuS@UW-21-016}
\title{UV/IR symmetries of the $S$-matrix and RG flow}
\author{\bf Silas R.~Beane}
\author{\bf and Roland C.~Farrell}

\affiliation{InQubator for Quantum Simulation (IQuS), Department of Physics,\\
  University of Washington, Seattle, WA 98195.}
\date{\mydate}

\abstract{The low energy $S$-matrix which describes non-relativistic
  scattering arising from finite-range forces has UV/IR symmetries
  that are hidden in the corresponding effective field theory (EFT)
  action. It is shown that the $S$-matrix symmetries are manifest as
  geometric symmetries of the RG flow of coupling constants in the EFT
  action in both three and two spatial dimensions. An
  example is given demonstrating that UV/IR symmetry breaking in the
  $S$-matrix implies strong constraints on the RG flow of the
  coefficients of the corresponding symmetry-breaking operators in the
  EFT.}  \notoc
\begin{document} \maketitle
\newpage

\section{Introduction}
\setcounter{page}{1}

\noindent Non-relativistic s-wave scattering with finite-range forces
exhibits two fixed points of the RG: the trivial fixed point
corresponding to no interaction, and the unitary fixed point where the
interaction strength takes the maximal value consistent with
unitarity~\cite{Weinberg:1991um} (for a review, see
Ref.~\cite{Braaten:2004rn}). It is not so well known that the
$S$-matrix has interesting properties with respect to a UV/IR
transformation which inverts the momentum~\cite{Beane:2020wjl}. Just
above threshold, where scattering is determined by the scattering
length, the momentum inversion maps the trivial RG fixed point into
the unitary RG fixed point and vice versa.  As a result, the UV/IR
transformation does not act simply on the scattering amplitude which
vanishes at the trivial fixed point.  Instead, it is the $S$-matrix,
which accounts for the trivial fixed point via the unit operator, that
manifests the UV/IR symmetry.  The UV/IR properties of the $S$-matrix
are not purely academic as they imply a conformal invariance which
allows the construction of simple geometrical models of scattering
phase shifts that have some remarkable
properties~\cite{Beane:2020wjl,Beane:2021B}. In addition, these novel
symmetries provide a new perspective on EFT descriptions of the
nucleon-nucleon (NN) interaction and facilitate the development of new
EFTs which enhance the convergence\footnote{For recent work which aims
  to improve the convergence of NN EFT see
  Refs.~\cite{SanchezSanchez:2017tws,Peng:2021pvo,Mishra:2021luw,Ebert:2021epn,Habashi:2020qgw,Habashi:2020ofb}.}
of the description at very low energies~\cite{Beane:2021C}.

The goal of this report is to show that the UV/IR properties of the
$S$-matrix are reflected in the (scheme dependent) beta-functions which characterize the
RG scale dependence of the EFT couplings. This will be explored for the non-relativistic EFT 
describing single-channel and two-channel scattering in both two and three spatial dimensions. 
This is applicable to the low energy NN interaction which will be used throughout
as the paradigmatic example.  This paper is organized as follows.
Section~\ref{sec:smatth} introduces the $S$-matrix and its UV/IR
symmetries (whose details are treated in appendix~\ref{app:Ward}). In section~\ref{sec:eftrg}, the EFT which matches to
the $S$-matrix in the scattering length approximation is reviewed and the
UV/IR symmetry is shown to be present in the RG flow of the
coefficients of momentum-independent local operators.
Constraints on higher-order EFT operators which follow from considering
UV/IR symmetry breaking in the $S$-matrix are considered in section~\ref{sec:eftsb}.
Section~\ref{sec:conc} summarizes and concludes.

\section{$S$-matrix theory and UV/IR symmetry}
\label{sec:smatth}

\noindent At very-low energies, the $S$-matrix which describes the
scattering of two species of equal-mass, spin-$1/2$ fermions
(labeled as nucleons and collectively denoted N), can be expressed
as~\cite{Beane:2018oxh}
\begin{eqnarray}
  \hat {\bf S} & = & \frac{1}{2}\left( S_1+ S_0 \right) \;\hat {\bf 1} \;+\; \frac{1}{2}\left( S_1- S_0 \right)\;\hat{{\cal P}}_{12} \ ,
   \label{eq:Sdefres}
\end{eqnarray}
where the $S$-matrix elements are 
\begin{eqnarray}
S_s &=& e^{2i\delta_s(p)} \ .
   \label{eq:Selementdef}
\end{eqnarray}
The $\delta_s$ are NN s-wave phase shifts with $s=0$ corresponding to the spin-singlet
($\si$) channel and $s=1$ corresponding to the spin-triplet ($\siii$)
channel, and $p$ is the center-of-mass momentum.
The SWAP operator is defined as
\begin{eqnarray}
\hat{{\cal P}}_{12} = \oneht \left({ \hat   {\bf 1}}+  \hat  {\bm \sigma} \cdot   \hat  {\bm \sigma} \right) \ ,
  \label{eq:swap}
\end{eqnarray}
where, in the direct-product space of the nucleon spins,
\begin{eqnarray}
 \hat {\bf 1} \equiv \hat {\cal I}_2\otimes  \hat {\cal I}_2 \ \ \ , \ \
  \hat {\bm \sigma} \cdot \hat {\bm \sigma} \equiv \sum\limits_{\alpha=1}^3 \
\hat{ { \sigma}}^\alpha \otimes \hat{{ \sigma}}^\alpha \ ,
  \label{eq:Hil}
\end{eqnarray}
with ${\cal I}_2$ the $2\times 2$ unit matrix and $\hat{ { \sigma}}^\alpha$ the Pauli matrices.
Very near threshold, the phase shifts are well approximated
by the $S$-matrix elements given at leading order in the effective range expansion (ERE)
\begin{eqnarray}
S_s &=& \frac{1- i a_s p}{1+ i a_s p} \ ,
   \label{eq:Selementscatt}
\end{eqnarray}
where the $a_s$ are the scattering lengths.

The fixed points of the RG are determined by the flow
of coupling constants with a change of scale in the EFT which gives rise to Eq.~(\ref{eq:Sdefres}).
In non-relativistic scattering, the $S$-matrix takes special
constant values at these fixed points~\cite{Beane:2018oxh}.  For a given
channel, the fixed points of the RG occur when the phase shifts,
$\delta_0$ and $\delta_1$, vanish (trivial fixed point) or are at ${\pi}/2$
(unitary fixed point); i.e. when $S_s=\pm 1$. Therefore the fixed
points of the full $S$-matrix occur when the phase shifts both vanish,
$\delta_1=\delta_0=0$, both are at unitarity, $\delta_1=\delta_0={\pi}/2$, or
when $\delta_1=0$, $\delta_0={\pi}/2$ or $\delta_1={\pi}/2$, $\delta_0=0$. The
$S$-matrices at these four fixed points are $\pm \hat {\bf 1}$ and
$\pm \hat{{\cal P}}_{12}$.  In the ERE, the fixed
points of the RG are reached at $a_1= a_0=0$, $|a_1|=| a_0|=\infty$,
and at $a_1=0\,,\, |a_0|=\infty$, $|a_1|=\infty\,,\, a_0=0$, with
all effective range and shape parameters taken to be vanishing (the
scattering length approximation).  If all inelastic thresholds are
absent, and $p$ is defined on the interval $[0,\infty)$, then all four
  of the RG fixed points are accessible (in a limiting sense) via
  scattering. In the scattering length approximation this can be seen
  to be a consequence of the invariance of the $S$-matrix with respect
  to the scaling transformation
\begin{eqnarray}
p\mapsto e^\beta p \ \ \ , \ \ \ a_s\mapsto e^{-\beta} a_s \ ,
 \label{eq:dilinv}
\end{eqnarray}
with $\beta$ an arbitrary parameter. Operationally, keeping $a$ fixed and scaling $p$ with $\beta$
positive (negative) accesses the $S$-matrix of large (small)
scattering lengths.  Hence, with $a_0$ and $a_1$ finite, the $S$-matrix
is a trajectory from $\hat {\bf 1}$ to $-\hat {\bf 1}$. With $a_0$
finite (at unitarity) and $a_1$ at unitarity (finite), the trajectory
originates at $-\hat{{\cal P}}_{12}$ ($\hat{{\cal P}}_{12}$) and again ends at
$-\hat {\bf 1}$.

In addition to the scale invariance of Eq.~(\ref{eq:dilinv}), the individual $S$-matrix
elements, $S_s$, transform simply with respect to the momentum-inversion transformation
\begin{eqnarray}
p\mapsto {1 \over a_s^2\, p} \ .
   \label{eq:singlechanmominv}
\end{eqnarray}
As this transformation maps threshold to asymptotic infinity and vice-versa, it is a UV/IR transformation. (This transformation and its associated Ward identity are considered in detail in appendix~\ref{app:Ward}.) As a transformation on phase shifts, one finds that
Eq.~(\ref{eq:singlechanmominv}) implies
\begin{eqnarray}
  \delta_s(p) \mapsto -\delta_s(p) \pm {\pi}/2 \ \ &,& \ \ S_s \to -S^*_s  \ \ ,
\end{eqnarray}
where the sign of the shift by $\pi/2$ is determined by the sign of
the scattering length.  Therefore, considering scattering near
threshold, this momentum inversion transformation interchanges the trivial and unitary fixed points of the RG. It will
be seen below how this transformation is manifest in the
running coupling constants of the EFT.

Consider the transformation on the momentum of the full $S$-matrix
\begin{eqnarray}
p\mapsto \frac{1}{|a_1 a_0| p} \ .
   \label{eq:moebius}
\end{eqnarray}
It corresponds to the following transformations on the phase shifts:
\begin{eqnarray} 
&&   \delta_0(p) \mapsto \delta_1(p) \pm \pi/2 \ \ \ , \ \ \ \ \ \, \delta_1(p) \mapsto \delta_0(p) \mp \pi/2 \ \ \ \ \ \ \ \; ({a_1 a_0 <0}) \ , \nn \\
  &&  \delta_0(p) \mapsto -\delta_1(p) \pm \pi/2 \ \ \ , \ \ \ \delta_1(p) \mapsto -\delta_0(p) \pm \pi/2 \ \ \ \ \ \ ({a_1 a_0 >0}) \ .
  \label{eq:confmoebiusiso2}
\end{eqnarray}
where the sign of the shift by $\pi/2$ is determined by the signs of the scattering lengths.
This is a conformal symmetry, in the sense that it
leaves the combination of phase shifts $\delta_0+\delta_1$ ($a_1 a_0
<0$) or $\delta_0-\delta_1$ ($a_1 a_0 >0$) invariant. This symmetry allows
the construction of simple geometrical models of scattering that are
investigated in Ref.~\cite{Beane:2020wjl,Beane:2021B}. 

The transformation of Eq.~(\ref{eq:moebius}) acts on the full $S$-matrix as
\begin{equation}
 \hat{{\bf S}} \to  e^{-2 i \Phi} \hat{{\bf S}}  \ \  \ (a_1 a_0 >0) \ \ \ , \ \ \ \hat{{\bf S}} \to  e^{2 i \Phi} \hat{{\bf S}}^*  \ \  \ (a_1 a_0 < 0) \ ,
\end{equation}
where $\Phi = \delta_0 + \delta_1 + \pi/2$ is a global phase. 
The consequences
of the scattering event for spin observables are encoded in the density matrix of the ``out'' state:
\begin{equation}
    \rho = \lvert \text{out} \rangle \langle \text{out} \rvert = \hat {\bf S} \lvert \text{in} \rangle \langle \text{in} \rvert \hat {\bf S}^{\dagger}
\end{equation}
which is unchanged by global phases of the $S$-matrix. As a result, for $(a_1 a_0 >0)$,
the UV/IR transformation leaves $\rho$ unchanged and generates a symmetry of all spin
observables measured on the ``out'' state.  For $(a_1 a_0 <0)$, the density matrix gets
mapped to the one which would be obtained if the signs of both scattering lengths were
flipped~\cite{Beane:2021B}.
As was shown in Ref.~\cite{Beane:2020wjl}, these simple transformations of the
density matrix imply that the spin-entanglement generated by the interaction is
invariant under the UV/IR transformation.

\section{Effective field theory and RG flow}
\label{sec:eftrg}

\subsection*{EFT action and potential}

\noindent It is interesting to ask whether the UV/IR properties of the
$S$-matrix are reflected in the EFT. The $S$-matrix in the
scattering-length approximation can be derived from an EFT with
highly-singular, momentum-independent, contact interactions. These
give rise to the quantum mechanical potential that enters the
Lippmann-Schwinger equation, which in turn generates the $S$-matrix.
The $S$-matrix models with exact UV/IR symmetry are,
by definition, UV-complete. Therefore one might expect the UV/IR
symmetry to be reflected in the flow of the EFT potential between
fixed points of the RG.

The EFT which describes low-energy s-wave scattering of nucleons is
constrained by spin, isospin and Galilean invariance (and, for the
case we consider, parity and time-reversal invariance). The
leading-order (LO) interactions in the Lagrangian density
are~\cite{Weinberg:1991um,Weinberg:1990rz}
\begin{equation}
{\cal L}_{\rm LO}
=
-\frac{1}{2} C_S (N^\dagger N)^2
-\frac{1}{2} C_T \left(N^\dagger{\bm \sigma}N\right)\cdot \left(N^\dagger{\bm \sigma}N\right) \ ,
\label{eq:interaction}
\end{equation}
where the field $N$ represents both spin states of the proton and neutron fields.  These interactions
can be re-expressed as contact interactions in the $\si$ and $\siii$
channels with couplings $C_0 = ( C_S-3 C_T) $ and $C_1 = (C_S+C_T)$
respectively, where the two couplings are fit to reproduce the $\si$
and $\siii$ scattering lengths.  The quantum-mechanical potential is
scheme dependent and can be read off from the Lagrangian density~\cite{Beane:2018oxh}
\begin{eqnarray}
{ V}(\mu)_\sigma
  & = &
\frac{1}{2}\left( { C}_{1}(\mu) +  { C}_{0}(\mu)  \right)  {\hat   {\bf 1}} +
\frac{1}{2}\left( { C}_{1}(\mu) -  { C}_{0}(\mu) \right) \hat{{\cal P}}_{12}
\ ,
   \label{eq:Vrescale}
\end{eqnarray}
where the $S$-matrix basis has been chosen.  In what follows
the flow of the potential with the RG scale, $\mu$, will be considered
in three and two spatial dimensions ($d=3,2$). 

\subsection*{RG flow in $d=3$}

\begin{figure}[!ht]
\centering
\includegraphics[width = 0.73\textwidth]{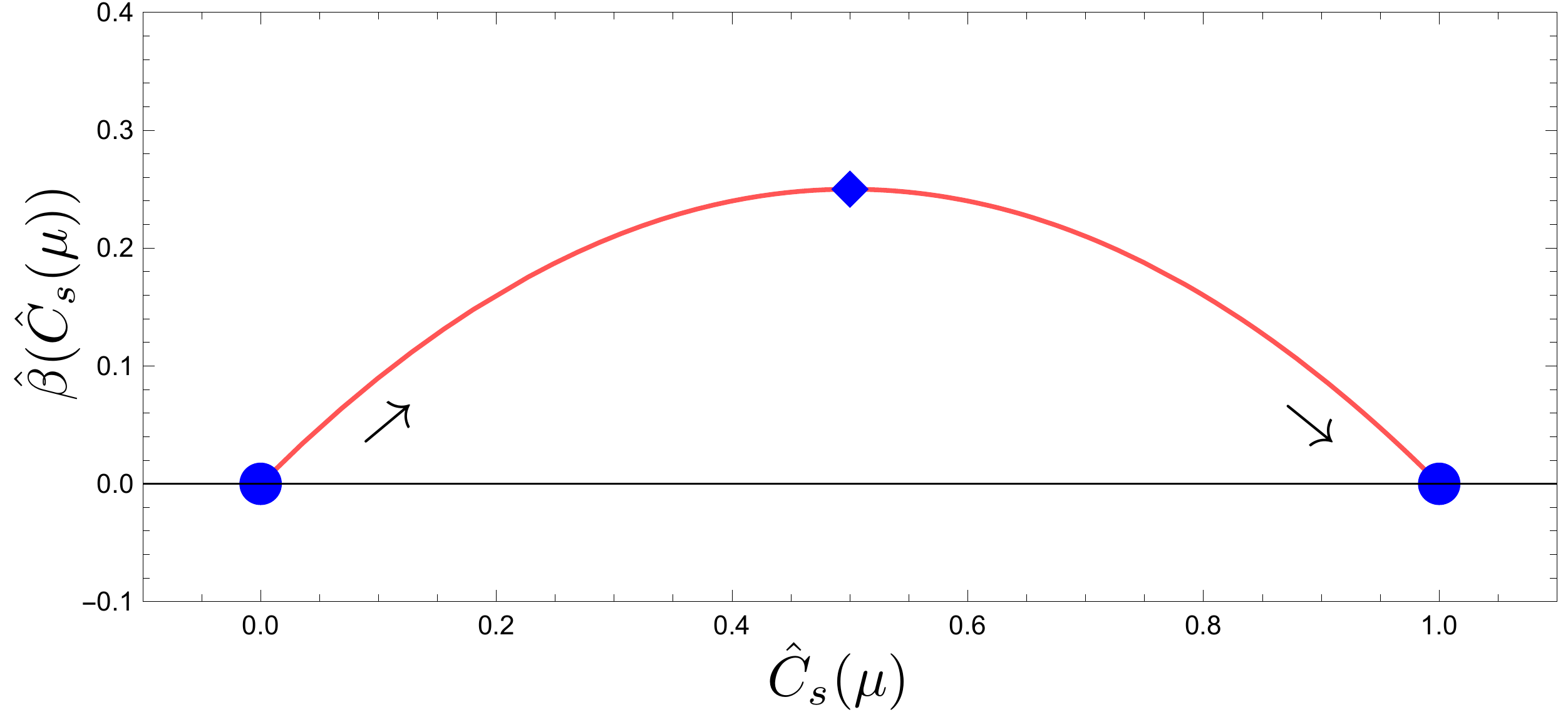}
\caption{The beta-function of Eq.~(\ref{eq:c0betafn}) plotted as a function of ${\hat C}_s(\mu)$. The blue dots are the RG fixed points.
  The beta-function curve evolves from $\mu=0$ (the trivial fixed point) to $\mu=\infty$ (the unitary fixed point). The diamond is the
fixed point of the scale-inversion symmetry, Eq.~(\ref{eq:singlechancutoffinv}), at ${\hat
  C}_s(\mu^\circ) = 1/2$.}
  \label{fig:eftc0beta}
\end{figure}

\noindent Solving the Lippmann-Schwinger equation with the potential
of Eq.~(\ref{eq:Vrescale}), or alternatively, summing to all orders
the loop diagrams with insertions of the operators in
Eq.~(\ref{eq:interaction}), leads to the $d=3$ NN scattering amplitude.
In dimensional regularization (dim reg) with the power-divergence subtraction (PDS)
scheme~\cite{Kaplan:1998tg,Kaplan:1998we} the amplitude is
\begin{equation}
  i \mathcal{A} = \frac{-i V(\mu)}{1 + M V(\mu) \left( \mu + i p\right) /4\pi} \ \ \ ,
   \label{eq:AitoV}
\end{equation}
where $M$ is the nucleon mass and $V(\mu)$ is the projection of
${V}(\mu)_\sigma$ onto a particular scattering channel. The PDS scheme
offers a clean way of accounting for the linear
divergences which appear in loops. The PDS couplings also exhibit the trivial and unitary RG fixed points
and, for $\mu \sim p$, scale so as to justify their non-perturbative treatment (power counting is manifest).  
The relation between the $\mu$-dependent coefficients
and the phase shifts in the scattering-length
approximation follows from matching the scattering amplitude to the ERE and is
\begin{equation}
  p \cot \delta_s =  - \left( \frac{4 \pi}{M C_s} + \mu \right) = -\frac{1}{a_s}\ .
\end{equation}
Therefore, the running couplings in the PDS scheme are
\begin{equation}
  C_s(\mu) \ =\ \frac{4 \pi}{M} \frac{1}{{1}/{a_s}-{\mu}} \ .
  \label{eq:C0PDS}
\end{equation}
There is a fixed point at $C_s=0$, corresponding to free particles ($a_s=0$),
and a fixed point at $C_s=C_\star$ corresponding to a divergent scattering
length (unitarity). It is convenient to rescale the couplings to 
${\hat C}_s \equiv C_s/C_\star$. The beta-functions for the rescaled couplings
are then 
\begin{equation}
  {\hat\beta({\hat C}_s)}\ =\  \mu \frac{d}{d\mu} {\hat C}_s(\mu) \ =\ -{\hat C}_s(\mu)\left({\hat C}_s(\mu)-1\right) \ ,
   \label{eq:c0betafn}
\end{equation}
which has fixed points at ${\hat C}_s=0$ and $1$, as shown in Fig.~(\ref{fig:eftc0beta}). The coupling is near the
trivial fixed point for $\mu < 1/|a_s|$, and near the non-trivial fixed point for $\mu
> 1/ |a_s|$. The four fixed points in the NN system are at ${\hat C}_1= {\hat C}_0=0$, ${\hat C}_1={\hat C}_0=1$,
and at ${\hat C}_1=0\,,\,{\hat C}_0=1$, and ${\hat C}_1=1\,,\, {\hat C}_0=0$.
In the space of rescaled couplings, these four points,
$(0,0)$, $(1,1)$, $(0,1)$ and $(1,0)$ furnish a representation of the Klein
group~\cite{Beane:2018oxh}.

Now recall that the momentum inversion, $p \to {1 /(a_s^2 p)}$, has the
effect of interchanging the trivial and unitary fixed points of
the $S$-matrix elements in the scattering length approximation.
In the EFT, under an inversion of the PDS RG scale
\begin{eqnarray}
\mu \mapsto {1 \over a_s^2 \, \mu} \ ,
   \label{eq:singlechancutoffinv}
\end{eqnarray}
the couplings transform as
\begin{equation}
   {\hat C}_s(\mu) \mapsto 1 -{\hat C}_s(\mu)\ .
\end{equation}
This implies that the beta-function evaluated at two scales related by an inversion are
equal, i.e. ${\hat\beta({\hat C}_s)}\rvert_{\bar{\mu}} = {\hat\beta({\hat
C}_s)}\rvert_{1/(a_s^2 \bar{\mu})}$, for any $\bar{\mu}$. This scale-inversion
transformation maps the two RG fixed points to one another and the UV/IR
transformation property of the phase shifts is reflected in the
$\mu$ dependence of the coupling. In addition, the beta-function is reflection
symmetric about the fixed point of the inversion transformation, ${\hat
  C}_s(\mu^\circ) = 1/2$, which occurs at
$\mu^\circ=|a_s|^{-1}$, as shown in Fig.~(\ref{fig:eftc0beta}).

The analysis presented above holds for other renormalization schemes as well, provided that they preserve the UV/IR symmetry; i.e. reproduce the scattering length approximation.
For example, consider regulating with a hard-cutoff in momentum space. 
The scattering amplitude is
\begin{eqnarray} 
\label{eq:cutoffamp}
   {\cal A} = {- C_s(\Lambda) \over 1 + M C_s(\Lambda) \Big[\Lambda +  i p  \tan^{-1}\left(i \frac{\Lambda}{p} \right) \Big]/(2\pi^2)}
\end{eqnarray}
where the cutoff-dependent coupling is defined as
\begin{eqnarray}
   C_s(\Lambda) = \frac{4\pi}{M} \frac1{1/a_s - 2\Lambda/\pi}
   \label{eq:Cscutoff} \ .
\end{eqnarray}
For finite $\Lambda>p$, the expanded $\tan^{-1}{ (i \Lambda / p)} $ term generates cutoff-dependent contributions to the amplitude to all orders in the effective range expansion~\cite{vanKolck:1998bw,Birse:1998dk}. These higher-order
terms break the UV/IR symmetry and therefore, to preserve the symmetry, higher-dimensional operators must be added to the EFT action to cancel these symmetry-breaking effect. For instance, expanding Eq.~\ref{eq:cutoffamp} gives
\begin{eqnarray}
\mathcal{A} &=& -\frac{4 \pi}{M} \frac{1}{1/a_s + i p}\left (1 +\frac{2}{\pi\Lambda} \frac{1}{1/a_s + i p}p^2 +  \mathcal{O}(\Lambda^{-2}) \right)
\end{eqnarray}
which evidently requires a momentum dependent counterterm ---a shift in the  $C_{2 \, s}$ operator that appears at NLO in the EFT expansion--- that scales like $\mathcal{O}(\Lambda^{-1})$~\cite{Kaplan:1998tg,Kaplan:1998we,vanKolck:1998bw}. In addition, insertions of this counterterm in perturbation theory will generate new terms in the amplitude that scale like positive powers of the cutoff, and whose removal will in turn require even higher dimensional counterterms~\cite{vanKolck:1998bw}.
This procedure of choosing counterterms to reproduce the scattering length approximation is nothing new; the identical procedure must be carried out in order to renormalize the cutoff EFT in a manner that preserves the Schr\"odinger symmetry Ward identities~\cite{Mehen:1999nd}, in the unitary, $\lvert a_s \rvert  \to \infty$, limit.

A key observation is that the cutoff dependence of $C_s (\Lambda)$ does not change as more counterterms are added, and the RG flow of this coupling in the cutoff scheme is the same as in PDS, Eq.~(\ref{eq:C0PDS}), but with $\mu \mapsto \frac{2}{\pi}\Lambda$. Therefore, the RG scale-inversion symmetry of the leading-order beta function
is not an artifact of a particular scheme, but rather a manifestation of
a physical property of the system which is reflected in the RG evolution
of the EFT couplings. Note that after all the higher-dimensional symmetry-restoring operators have been added, the theory is valid for all momenta and varying the arbitrary cutoff sets the scale for the physical process. 

\begin{figure}[!ht]
\centering
\includegraphics[width = 0.8\textwidth]{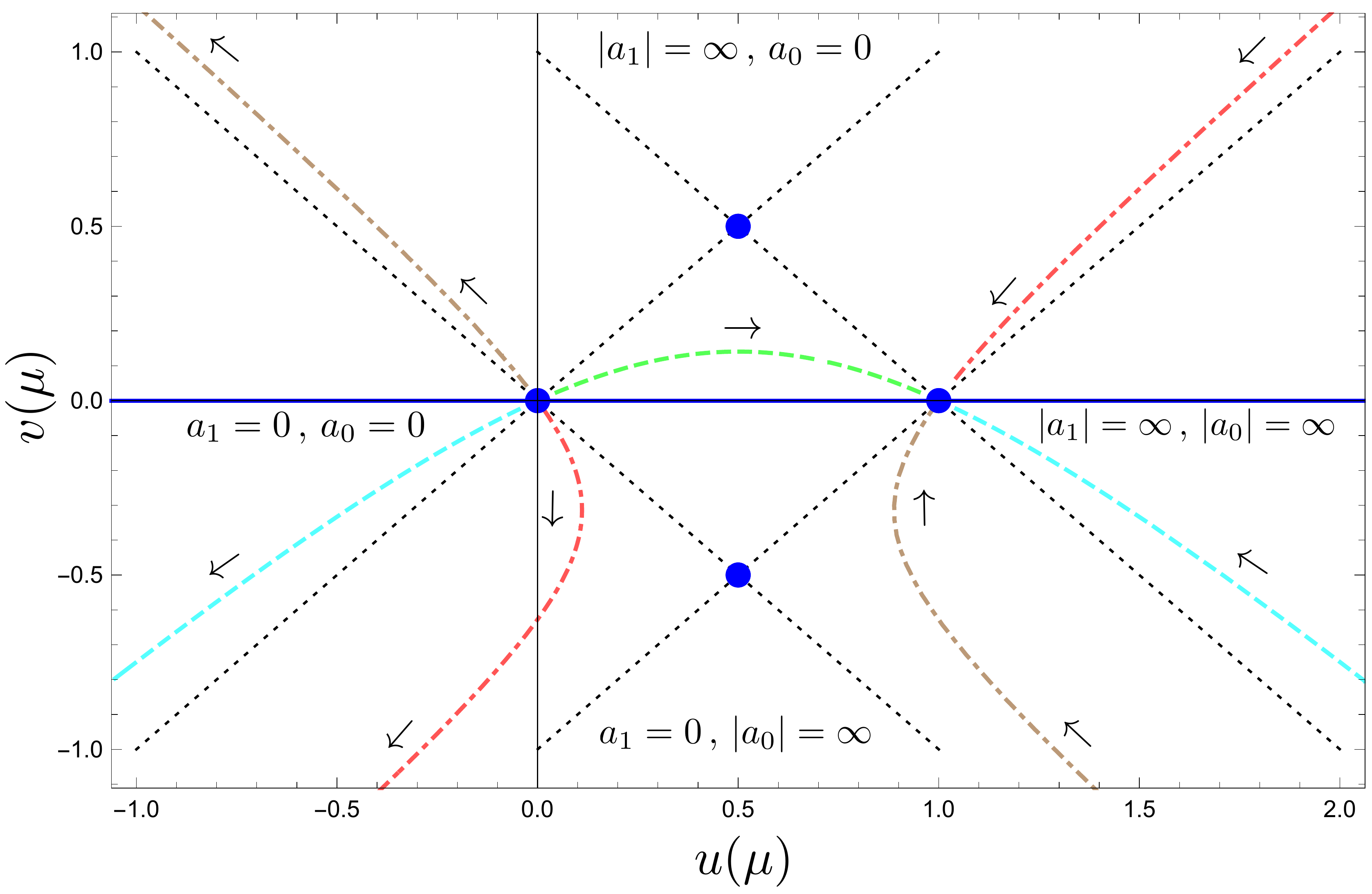}
\caption{The four fixed points of the RG in the $u-v$ plane in $d=3$. The
red-dashed line is the RG trajectory corresponding to the physical NN scattering lengths. The brown-dashed lines flip the signs of the physical scattering lengths.
The green (cyan) dashed curves are RG trajectories that exhibit the UV/IR symmetry and correspond to
both scattering lengths negative (positive). The arrows indicate direction of increasing $\mu$. Note that part of the cyan curve is not shown.}
  \label{fig:eftrhom}
\end{figure}

Returning to two-channel s-wave scattering, it is convenient to define the components of the re-scaled potential as
\begin{eqnarray}
u(\mu)  \; = \;  \frac{1}{2}\left( {\hat C}_{1}(\mu) +  {\hat C}_{0}(\mu)  \right) \ \ , \ \
v(\mu) \;=\; \frac{1}{2}\left( {\hat C}_{1}(\mu) -  {\hat C}_{0}(\mu) \right)  \ .
\label{eq:uvCs}
\end{eqnarray}
In the $u-v$ basis the RG fixed points of the rescaled potential are at $(0,0)$, $(1,0)$, $(1/2,1/2)$ and $(1/2,-1/2)$,
and the components of the potential flow with the RG according to the algebraic curve
\begin{eqnarray}
v\left(v-{\bar w}\right) \ = \ u\left( u-1\right) \ \ \ , \ \ \ {\bar w} \ \equiv \ \frac{a_1 + a_0}{a_1 - a_0}  \ .
   \label{eq:EFTconic}
\end{eqnarray}
The curves for all combinations of signs of scattering lengths are plotted in Fig.~(\ref{fig:eftrhom}). 

At the end of section~\ref{sec:smatth} it was pointed out that when $a_1 a_0 > 0$ the momentum inversion symmetry is an exact symmetry of the
density matrix. In the figure this corresponds to the green and cyan curves which have a reflection symmetry about
the line $u = 1/2$. When $a_1 a_0 < 0$ the momentum inversion transformation maps the density matrix into one obtained from flipping the signs of the
scattering lengths. In the EFT this is seen through the
reflection about the line $u = 1/2$ mapping the trajectories with $a_0 > 0$ and $a_1 < 0$ (brown curve) and $a_0 < 0$ and $a_1 > 0$ (red curve) into each other.
In either case the reflection is generated by the scale-inversion 
\begin{eqnarray}
\mu \mapsto \frac{1}{\lvert a_1 a_0 \rvert \mu} \ .
   \label{eq:moebius2}
\end{eqnarray}
Therefore, in the EFT, the symmetry properties of the density matrix are encoded in
a geometric, reflection symmetry of the RG flow of the coupling constants. It
is curious that when the system is unbound in both s-wave channels, the RG flow is confined to the rhombus
formed by the four RG fixed points in much the same way that the $S$-matrix is confined via unitarity to the
flat torus defined by the two s-wave phase shifts~\cite{Beane:2020wjl}.

\subsection*{RG flow in $d=2$}

\noindent In order to further investigate the relation between UV/IR
symmetries of the $S$-matrix and RG flow, consider deforming the
scattering system to $d=2$ via an anisotropic harmonic
trap~\cite{PhysRevA.64.012706} or by compactifying a
dimension~\cite{PhysRevA.93.063631}. The $S$-matrix element in the
$d=2$ scattering length approximation becomes
\begin{eqnarray}
S_s &=& \frac{\log{\left ( {\textit{\textbf a}}_s^2 p^2 \right )} + i \pi}{\log{\left ( {\textit{\textbf a}}_s^2 p^2 \right )} - i \pi}  \ ,
   \label{eq:Selementscatt2d}
\end{eqnarray}
where ${\textit{\textbf a}}_s$ is the intrinsically positive $d=2$ scattering length.
Here the momentum inversion transformation, $p \mapsto 1/({\textit{\textbf a}}_s^2 p)$, maps $S_s\mapsto S_s^*$.

The scattering amplitude obtained in the EFT has a logarithmic divergence which requires regularization and
renormalization. Using dim reg with the $\overline{\text{MS}}$ scheme one finds that the couplings run
with the RG scale as~\cite{Kaplan:2005es,Beane:2010ny}
\begin{equation}
    C_s(\mu) = \frac{- 4 \pi}{M \log{\left ( {\textit{\textbf a}}_s^2 \mu^2 \right )}} \ .
\end{equation}
Notice that the coupling has a pole at $\mu = {\textit{\textbf a}}_s^{-1}$ where it changes sign. With $\mu > {\textit{\textbf a}}_s^{-1}$, 
$C_s(\mu)<0$ corresponding to attraction, and the coupling appears asymptotically free; i.e. flows to zero logarithmically.
With $\mu < {\textit{\textbf a}}_s^{-1}$,  $C_s(\mu)>0$ corresponding to repulsion, and the coupling runs into a Landau pole
in the UV at $\mu = {\textit{\textbf a}}_s^{-1}$.

\begin{figure}[!ht]
\centering
\includegraphics[width = 0.73\textwidth]{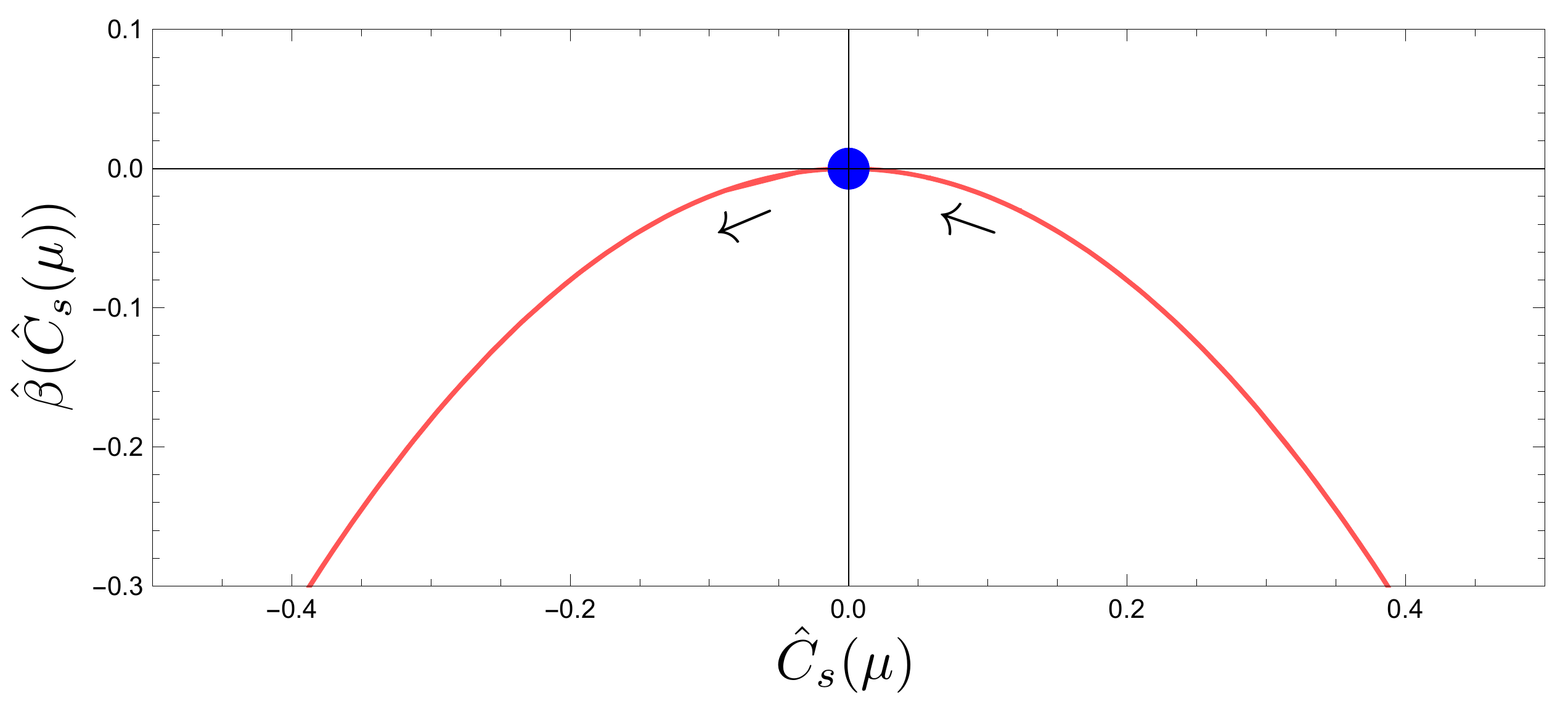}
\caption{The beta-function of Eq.~(\ref{eq:c0betafn2d}) plotted as a function of ${\hat C}_s(\mu)$. The blue dot is the RG fixed point.
  The beta-function curve evolves from $\mu=0$ (the trivial fixed point) to $\mu=\infty$ (again the trivial fixed point). The
fixed point of the scale-inversion symmetry, $\mu^\circ={\textit{\textbf a}}_s^{-1}$,  is at asymptotic infinity.}
  \label{fig:eftc0beta2d}
\end{figure}

It is convenient to define the dimensionless couplings, $\hat{C}_s(\mu) = {- M}C_s(\mu)/({4 \pi})$, whose beta-functions are
\begin{equation}
{\hat\beta({\hat C}_s)}\ =\    \mu \frac{d}{d\mu} \hat{C}_s(\mu) = -2 \hat{C}^2_s(\mu) \ .
  \label{eq:c0betafn2d}
\end{equation}
There is a single RG fixed point $\hat{C}_s = 0$, which is reached asymptotically at $\mu = 0$ and at $\mu = \infty$ as seen in
Fig.~(\ref{fig:eftc0beta2d}). Under an inversion of the RG scale,
$\mu \mapsto ({\textit{\textbf a}}_s^2 \mu)^{-1}$, the running couplings transform as
\begin{equation}
    \hat{C}_s(\mu) \mapsto -\hat{C}_s(\mu) \ ,
\end{equation}
which implies ${\hat\beta({\hat C}_s)}\rvert_{\bar{\mu}} = {\hat\beta({\hat
C}_s)}\rvert_{1/(a_s^2 \bar{\mu})}$ for any $\bar{\mu}$. The fixed point of the scale-inversion
transformation is at the Landau pole, $\mu^\circ = {\textit{\textbf a}}_s^{-1}$.

\begin{figure}[!ht]
\centering
\includegraphics[width = 0.7\textwidth]{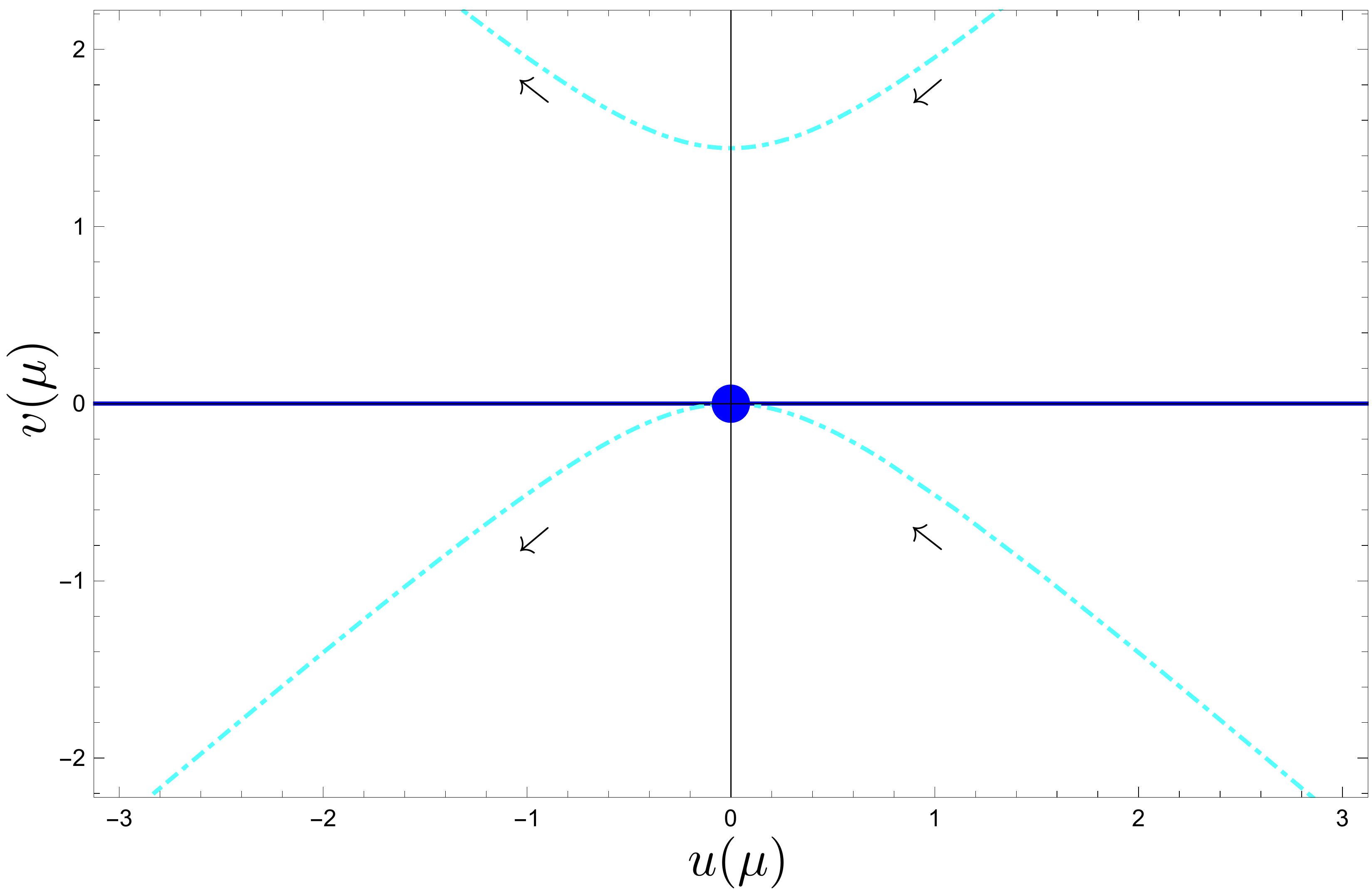}
\caption{The single fixed point of the RG in the $u-v$ plane in $d=2$.
The cyan dashed curve is the RG trajectory that exhibits the UV/IR symmetry. The trajectory begins and ends at the RG fixed point.  The arrows indicate direction of increasing $\mu$.}
  \label{fig:2dRG}
\end{figure}

Considering both s-wave scattering channels simultaneously, the momentum inversion
transformation of the $S$-matrix generalizes to $p \mapsto ({\textit{\textbf
    a}}_1 {\textit{\textbf a}}_0 \, p)^{-1}$ and leaves the
    density matrix invariant. Defining $u(\mu)$ and $v(\mu)$ as in $d = 3$, the components
of the potential flow with the RG according to the algebraic curve
\begin{eqnarray}
{\bar w}\left( v^2-u^2\right) = v\ \ \ , \ \ \ {\bar w} \ \equiv \  \log \left(\frac{ {\textit{\textbf a}}_1}{{\textit{\textbf a}}_0}   \right) \ .
   \label{eq:EFTconic2d}
\end{eqnarray}
The UV/IR transformation on the potential, via the inversion of
the scale, $\mu \mapsto ({\textit{\textbf a}}_1 {\textit{\textbf a}}_0 \,
\mu)^{-1}$, generates a reflection about the $v$-axis, $u \mapsto
-u$, $v \mapsto v$, in the $u-v$ plane. The fixed point of the scale-inversion
symmetry occurs at $\mu^{\circ} = ({\textit{\textbf a}}_1 {\textit{\textbf a}}_0)^{-1/2}$ where $\hat{C}_0 = -\hat{C}_1$. The RG
trajectory of the
potential is shown in Fig.~(\ref{fig:2dRG}). It is clear that, as in
three spatial dimensions, a symmetry of the density matrix is encoded in the
EFT as a reflection symmetry of the RG flow of the potential.
\section{UV/IR symmetry breaking}
\label{sec:eftsb}

\noindent The correspondence between the UV/IR transformation
properties of the S-matrix and the RG flow of EFT couplings is not
confined to the scattering length approximation. The UV/IR
transformation properties of the momentum-dependent corrections in the
effective range expansion are reflected in the EFT interaction
potential via constraints on the RG flow of the associated EFT
couplings.  This section will focus on corrections to the
single-channel example in $d=3$ considered above: the case of a large
s-wave scattering length with perturbative effective range
corrections~\cite{Kaplan:1998tg,Kaplan:1998we,vanKolck:1998bw,Birse:1998dk}.
The methods used in this section are similar to those used in
Ref.~\cite{Beane:2021C} where a distinct UV/IR symmetry was imposed to
constrain the EFT relevant for a non-perturbative treatment of the
effective range.  It will be shown that the UV/IR transformation
properties of the scattering-amplitude corrections reproduces key
features of the known RG flow of higher-order couplings in the EFT,
without explicit calculation of the higher-order effects.

Treating the scattering length to all orders with effective range and
shape parameter effects treated perturbatively, the $d=3$
s-wave\footnote{Note that in this section all spin labels are dropped
  and the results apply equally to the spin-singlet and spin-triplet
  channels.} scattering amplitude is given by the ERE which, up to
NLO,
is~\cite{Kaplan:1998tg,Kaplan:1998we,vanKolck:1998bw,Birse:1998dk}
\begin{eqnarray}
\mathcal{A} &=& -\frac{4 \pi}{M} \frac{1}{1/a + i p}\left (1+ \frac{r/2}{1/a + i p}p^2\right ) \equiv  \mathcal{A}_{-1} (p) + \mathcal{A}_{0}(p) 
\label{eq:pertamp}
\end{eqnarray}
where $r$ is the effective range.  Defining $\hat{p} \equiv a p$, at each order the scattering amplitude transforms simply under the UV/IR transformation, $\hat{p} \to 1/\hat{p}$,
\begin{eqnarray}
\mathcal{A}_{-1}(1/\hat{p}) &=& -i \hat{p}\,\mathcal{A}_{-1}(\hat{p})^* \ \ ,\ \ \mathcal{A}_{0}(1/\hat{p}) \, =\, -\hat{p}^{-2}\,\mathcal{A}_{0}(\hat{p})^* \ .
\label{eq:UVIRamp}
\end{eqnarray}
The scattering amplitude is generated from an EFT of contact interactions via a renormalized, on-shell, tree-level interaction potential
\begin{equation}
    V = C_0(\hat{\mu}) + C_2(\hat{\mu}) \hat{p}^2/a^2 \equiv V_{-1}(\hat{\mu}) + V_{0}(\hat{\mu}, \hat{p})
    \label{eq:pertpot}
\end{equation}
where $C_n$ is the coefficient of the four-fermion interaction with
$n$ derivatives and $\hat{\mu} = a \mu$.  Note that here the RG scale
$\mu$ can represent the PDS scale or a hard cutoff\,\footnote{For the case of cutoff regularization, we are omitting from the potential the counterterms necessary to exactly match onto Eq.~(\ref{eq:pertamp}).}, and therefore
Eq.~(\ref{eq:AitoV}) is not assumed. The potential relevant for
momentum $\hat{p}$ is obtained by setting $\hat{\mu} = \hat{p}$ and,
in a particular renormalization scheme, the UV/IR properties of the
potential should mirror those of the scattering amplitude that it
generates, i.e.  Eq.~(\ref{eq:UVIRamp}). Hermiticity implies that $V =
V^*$ and that there will be no imaginary phases in the UV/IR
transformation. The assumption that the interaction, as represented by
the potential, reflects the UV/IR transformation properties of the
amplitude at each order in the EFT expansion, suggests that by
imposing the UV/IR transformation, $\hat{\mu} \to 1/\hat{\mu}$ and
$\hat{p} \to 1/\hat{p}$, and setting $\hat{\mu} = \hat{p}$, the
potential should transform as
\begin{eqnarray}
    V_{-1}(1/\hat{p}) &=& \epsilon_{\scriptscriptstyle -1}\, \hat{p}\,V_{-1}(\hat{p}) \ \ ,\ \
    V_{0}(1/\hat{p}, 1/\hat{p}) \,=\, \epsilon_{\scriptscriptstyle 0}\, \hat{p}^{-2}\, V_{0}(\hat{p}, \hat{p}) 
    \label{eq:UVIRpot}
\end{eqnarray}
where $(\epsilon_{\scriptscriptstyle -1,0})^2=1$. 
This in turn implies that the renormalized couplings transform as
\begin{eqnarray}
    C_0(1/\hat{\mu}) &=& \epsilon_{\scriptscriptstyle -1}\, \hat{\mu}\,C_0(\hat{\mu}) \ \ ,\ \
    C_2(1/\hat{\mu}) \,=\, \epsilon_{\scriptscriptstyle 0}\, \hat{\mu}^2\, C_2(\hat{\mu}) \ .
    \label{eq:UVIRCn}
\end{eqnarray}
Hence, once $C_0$ is determined, the UV/IR transformation properties imply
$C_2\propto r (C_0)^2$, as confirmed by explicit calculation of higher
order loop effects in the EFT using both PDS and cutoff\,\footnote{Using cutoff 
regularization, the insertion of the $C_n$ operators in the EFT
calculation generates increasingly singular and non-linear cutoff dependence. However,
these singular contributions are
canceled by existing counterterms~\cite{vanKolck:1998bw}, as noted in section~\ref{sec:eftrg}.}
regularization~\cite{Kaplan:1998tg,Kaplan:1998we,vanKolck:1998bw,Birse:1998dk}.
This readily extends to higher orders in the EFT expansion; it is easy
to check that the UV/IR transformations of shape parameter and
higher-order amplitudes constrains $C_{2n}$ to scale as specific
powers of $C_0$.  Furthermore, from Eq.~(\ref{eq:pertamp}), it is
clear that the couplings should all vanish as $a \to 0$. Assuming
polynomial dependence on $a$ and $\mu$, dimensional analysis then
implies that $C_0(\hat{\mu}) = a/M \,f(\hat{\mu})$, and one class of
solutions to the UV/IR constraint is given by
\begin{equation}
    f(\hat{\mu}) = -\frac{c\left(\hat{\mu}+1\right)\hat{\mu}^n}{\hat{\mu}^{2n+2}+\epsilon_{\scriptscriptstyle -1}}
\end{equation}
where $c$ and $n$ are real constants. Setting
$\epsilon_{\scriptscriptstyle -1}=-1$, $n=0$, and $c = 4\pi$ recovers $C_0(\mu)$ in the PDS scheme~\cite{Kaplan:1998tg,Kaplan:1998we}.

\section{Summary and Conclusion}
\label{sec:conc}

\noindent In both two and three spatial dimensions, the scattering
length approximation to the low-energy, s-wave $S$-matrix has a UV/IR
symmetry which leaves the density matrix of the ``out'' state
invariant. While this is not a symmetry of the scattering amplitude or
effective action in the sense of a transformation on the fields, the
UV/IR symmetry does appear as a symmetry of the RG evolution of the
EFT couplings. In a sense, the echo of the $S$-matrix symmetry in RG
evolution is a consistency check that indeed the $S$-matrix is
rendered UV complete by the symmetry. As the UV/IR transformation maps
threshold to asymptotic infinity and vice versa, its presence signals a
UV-complete description of the scattering event. That is, scattering
is well defined at all distance scales. Clearly then, the UV completeness
should be reflected in the interaction itself, which should also be
defined over all distance scales. In the EFT of contact operators,
this is reflected by the presence of the RG scale $\mu$ which can be
chosen to take any value, and in the UV/IR symmetry of the beta-function.

It was shown that the UV/IR symmetry has utility beyond the scattering
length approximation, where the simple RG evolution between two fixed
points breaks down. Indeed, the manner in which perturbations around
the scattering length approximation break the UV/IR symmetry strongly
constrains the RG flow of higher dimensional couplings in the
corresponding EFT. In addition, there is another class of UV/IR
symmetric $S$-matrices which include effective ranges that are
correlated to the scattering
lengths~\cite{Beane:2020wjl,Beane:2021B}. In that case, the symmetry
also appears in the quantum mechanical potential which gives rise to
the $S$-matrix, albeit in a different manner than was shown in this
paper~\cite{Beane:2021C}.

One of the important conclusions of this work is that there are
symmetries of a scattering process which are not manifest symmetries
of the scattering amplitude.  This arises because observables measured
on an ``out'' state depend on the full wave function after
scattering. That is, the contribution from the part of the wave
function which does not scatter (corresponding to the identity
operator in the $S$-matrix) is crucial in constructing the ``out''
state.  Furthermore, as the full wave function may decompose into many
scattering channels, each with their own scattering amplitude, there
can be symmetries which are only apparent if all scattering channels
are viewed holistically.  It would be interesting if such $S$-matrix
symmetries can arise in other contexts, perhaps unrelated to
momentum inversion.

One important issue that has not been addressed in this brief report
is the spacetime nature of the UV/IR symmetry. As the UV/IR
transformation is an inversion of momentum, it necessarily involves a
scale transformation. Given the results of appendix~\ref{app:Ward}, it appears promising to investigate whether the
UV/IR symmetry could be understood as an extension of Schr\"odinger
symmetry~\cite{Hagen:1972pd,Niederer:1972zz,Mehen:1999nd,Nishida:2007pj}
to systems with finite scattering length.

\appendix
\section{Momentum inversion Ward identity}
\label{app:Ward}
\noindent This appendix considers a generalization of the momentum inversion transformation of Eq.~(\ref{eq:singlechanmominv}) 
and derives the associated Ward identity. Consider the $S$-matrix of Eq.~(\ref{eq:Selementscatt}).
If one allows the momentum, $p$, to span the entire real line, then with respect to the real M\"obius transformation
\begin{eqnarray} \label{eq:mob1a}
p \mapsto {\vartheta\,p+ 1/a_s \over {\pm\left(a_s\,p-\vartheta \right)}} \ ,
   \end{eqnarray}
with $\vartheta$ an arbitrary real parameter, the $S$-matrix transforms as
\begin{eqnarray} \label{Smatmobius1a}
  S \mapsto \frac{\vartheta \pm i}{\vartheta\mp i}\,\begin{cases}
  S^* \ ,\\
  S\ .
\end{cases}
  \end{eqnarray}
That is, the $S$-matrix transforms to itself or its complex conjugate,
times a constant complex phase. Choosing the positive
sign in the transformation law, and $\vartheta=0$ recovers the UV/IR
transformation of Eq.~(\ref{eq:singlechanmominv}).

The M\"obius transformation is a general mapping of the momentum to itself and therefore
generally contains UV/IR transformations. In what follows, the Ward identity
for this symmetry will be derived. Let $\hat p\equiv a_s\, p$, and
consider the infinitesimal version of Eq.~(\ref{eq:mob1a}), with the
minus sign chosen in the transformation law. Since $\vartheta$ large
recovers the identity, take $\vartheta=1/\epsilon$ with $\epsilon$
infinitesimal. Then we see that Eq.~(\ref{eq:mob1a}) becomes
\begin{eqnarray} \label{eq:mob2a}
\hat p \mapsto {\hat p+ \epsilon \over {-\epsilon\,\hat p+1}}  \ .
   \end{eqnarray}
Now consider the infinitesimal translation
\begin{eqnarray} \label{eq:lm1a}
\hat p \mapsto  \hat p\;+\; \epsilon \ ,
\end{eqnarray}
the infinitesimal dilatation
\begin{eqnarray} \label{eq:l0a}
\hat p \mapsto  e^\epsilon \hat p = \hat p\;+\; \epsilon\,\hat p \;+\; \mathcal{O}(\epsilon^2) \ ,
\end{eqnarray}
and finally consider two inversions with a translation in between,
\begin{eqnarray} \label{eq:l1a}
\hat p \mapsto \left({\hat p}^{-1}-\epsilon\right)^{-1} = \frac{\hat p}{1-\epsilon\,\hat p} = \hat p\;+\; \epsilon\,\hat p^2 \;+\; \mathcal{O}(\epsilon^2) \ .
   \end{eqnarray}
This final step is critical in giving an infinitesimal description of momentum inversion.
These three transformations are generated by the differential operators $L_{-1}$, $L_{0}$,  and $L_{1}$, respectively, where
\begin{eqnarray} 
L_k\equiv - {\hat p}^{k+1}\frac{\partial}{\partial \hat p}  \ .
   \end{eqnarray}
These satisfy the ${\mathfrak sl}(2,\mathbb{R})$ algebra:
\begin{eqnarray} 
&& {[\, L_1\,  ,\, L_{-1}\, ]}\, =\, 2\,L_0 \ \ \ ,\ \ \  {[\, L_{\pm 1}\,  ,\, L_0\, ]}\, =\, \pm\,L_{\pm 1} \ .
\label{eq:sl2ralga}
\end{eqnarray}
Note that the general M\"obius transformation matrix of Eq.~(\ref{eq:mob1a}) has determinant $\mp(\vartheta^2+1)$
and therefore is an element of $PSL(2,\mathbb{R})$ only in the special case where the minus sign is chosen in the transformation law
and $\vartheta=0$\,\footnote{If $a_s = \pm 1$ then Eq.~(\ref{eq:mob1a}) is an element of the modular group, $PSL(2,\mathbb{Z})$.}. The general M\"obius transformation is an element of $PGL(2,\mathbb{R})$.

It is easy to check that the (infinitesimal) M\"obius transformation of Eq.~(\ref{eq:mob2a}) is constructed by 
successive transformations of Eq.~(\ref{eq:l1a}) and Eq.~(\ref{eq:lm1a}); that is, it is generated by $L_{1}$ and $L_{-1}$. Indeed the Ward identity is:
\begin{eqnarray} 
\left(L_{1} + L_{-1}-2 i \right)S\ =\ 0 \ .
   \end{eqnarray}
The general solution of this differential equation is Eq.~(\ref{eq:Selementscatt}), up to an overall complex coefficient.
Note that the (broken) on-shell dilatation Ward identity of the Schr\"odinger group~\cite{Mehen:1999nd} takes
the form~\cite{Beane:2020wjl}
\begin{eqnarray} 
  L_0 \: S&=& -\oneht\left( S^2-1\right)  \,
\end{eqnarray}
and, as expected, is respected at the RG fixed points, $S=\pm 1$.


\section*{Acknowledgments}

\noindent We would like to thank David B.~Kaplan and Martin J.~Savage for valuable discussions regarding this work. This work was supported by the U.~S.~Department of Energy
grants {\bf DE-FG02-97ER-41014} (UW Nuclear Theory) and {\bf DE-SC0020970}
(InQubator for Quantum Simulation).

\bibliographystyle{JHEP}
\bibliography{bibi}

\end{document}